# The Future of Artificial Intelligence and its Social, Economic and Ethical Consequences

Burhan Rashid Hussein
*Mathematic and Computing Sciences,
Universiti Brunei Darussalam*
Brunei Darussalam
burhr2@gmail.com

Chongomweru Halimu
*School of Computing and Informatics,
Universiti Teknologi Brunei*
Brunei Darussalam
halimkas@gmail.com

Muhammad Tariq Siddique
*School of Computing and Informatics,
Universiti Teknologi Brunei*
Brunei Darussalam
m.tariq.siddique@gmail.com

*Abstract*—Recent development in AI has enabled the expansion of its application to multiple domains. From medical treatment, gaming, manufacturing to daily business processes. A huge amount of money has been poured into AI research due to its exciting discoveries. Technology giants like Google, Facebook, Amazon, and Baidu are the driving forces in the field today. But the rapid growth and excitement that the technology offers obscure us from looking at the impact it brings on our society. This short paper gives a brief history of AI and summarizes various social, economic and ethical issues that are impacting our society today. We hope that this work will provide a useful starting point and perhaps reference for newcomers and stakeholders of the field.

*Keywords — Artificial intelligence, chatbot, ethics, AI bias*

## I. INTRODUCTION

Although Artificial intelligence (AI) came into existence in ancient times, only in 1956 the official name was formally given [1]. Since that time AI research has undergone a period of optimism and disappointment due to the slow progress which was observed. There was a fluctuation in progress made until after 1993. Research began to pick up again after that, and in 1997, IBM's Deep Blue became the first computer to beat a chess champion when it defeated Russian grandmaster Garry Kasparov [2]. This was the beginning of a new era in the AI field. In the last two decades, much has been done in academic research, but AI has been only recently recognized as a paradigm shift. A lot of progress is now being made as the investment has been going into the field. AI researches highly depends on funding since it's a long-term research field and requires an enormous amount of effort and resources.

AI techniques are being used to predicting deforestation before it happens, NASDAQ stock now monitors trades to see if insider trading is going on, NASA use AI methods to schedule payload operation, diagnosis of acute leukaemia, breast, and pancreatic cancer, and also predict patient survival with breast cancer, to car automation, and many other areas [3]. This application area highlights the importance and the benefits that AI technology brings us today. AI or machine intelligence can be devoted as an activity of making machine intelligent. That intelligence can be seen as the quality that enables an entity to function appropriately and with foresight in its environment [4]. But AI research defines AI as the study of intelligent agents, an agent that can perceive its environment and takes an action that maximizes the outcome or goal [5]. Advancements in computing power, theoretical understanding, and a large amount of data have enabled AI techniques to be an essential part of the technological revolution and in helping to solve many complex problems of our daily life. Take an example of recognizing object from images which just seemed like a sci-fi of AI a few years ago but now a simple convolutional neural network sits at the back and does the work [6]. All this may just be the begging of the new era of technological advancement. A huge amount of money has been poured into AI research due to its exciting discoveries and as the days go on it's getting better and better. But perhaps the capabilities that the AI brings obscure us from looking at the consequences that come with it. In this paper, we try to review the current social, economic, and ethical consequences that AI technologies will bring.

## II. SOCIAL AND ETHICAL CONSEQUENCES OF AI

Social consequences involve those effects which the technology will directly or indirectly impact our life from an individual perspective, community, and society at large. The integration of AI is now transforming our daily life inevitably [7]. We are using different products powered by AI without even noticing. Google Assistant, Amazon's Alexa, Roomba vacuum cleaner, chatbots, and many other products in our world today are mostly powered by AI technology. The popularity of the technology has been assisted by the increase in computational power and a sheer increase in the amount of data [8]. The excitement about AI and its future impact on health care, our economy, climatic changes, and the education system are promising. Despite AI progress, few ethical issues have been presented. With the technological approach to be human-like, the threats and unpredictable complication cannot be overseen [9].

## III. UNEMPLOYMENT

With the automation of processes dominating the AI field. Job displacement is one of the huge consequences that AI will have. In the past decade, human has relied on physical work and investing their time to earn. With the technology still rising and smart intelligent robots have been developed, we may see a future where individuals are being paid just for being citizens which will be important in helping to combat the job-stealing automation [10]. Despite the threat, a look of some automation like autonomous vehicles seems like a good ethic choice as it will help in a significant reduction of the accident if successfully implemented. One study conducted by Frey et al. found that 47% of total US employment is at risk of being replaced by machines over the next two decades [11]. AI-related jobs are now dramatically increasing in demand. According to the annual report AI index report [3], the AI skills requirements in job-listing have continued to grow multiples times from year to year. These statistics indicate the increasing need for AI skills to survive in the job market. But on the bright side, the replacement of machines could finally see humans doing what they like and other non-labour activities like family caring, community activities and found other new ways to help the society [12].

## IV. INEQUALITY

Another dramatic effect of AI technology is the widening of the wealth gap. As technology cuts off the human workforce in different companies, this means the revenue will go to fewer people. Hence AI-driven companies will make all the money while fewer people will benefit. Although some people argued that automation won't be the source of increased unemployment still it can destroy middle-range jobs while increasing those on the low and high-end jobs. This will augment social inequality and amplify the gap between low and higher-end job earnings [13]. The AI startups are now benefiting from the investments as the potential of the technology indicating its potential towards businesses. This brings the initiative for the society and their governments to create policies that will supervisor and ensure equal distribution of wealth among individuals to combat the increasing gap.

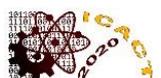

## V. HUMANITY

With the rise of AI bot, modern human interaction is also changing. A remarkable AI bot named Eugene Goostman won the Turing challenge for the first time [14]. In this challenge, the bot interacted with a human in a text input format and the user was required to guess whether they were chatting with a machine or human. The bot fooled more than half of the human raters into thinking they have been talking to their fellow human being. Unlike the human nature of being tired and changing moods, these machines are more efficient and have a wide environment and resources for making relationships with humans. The chatbot of today not only can answer a variety of human questions but also can establish an emotional connection with humans [6],[15].

We have already started to see the beneficial use of bots. One good example is the visabot which help a user better understand American immigration laws and apply for the right visa based on persons need. The chatbot helps in smoothing visa processes, save time, and ensure already filled in visa forms[1]. Another example is a Microsoft chatbot which was created to interact with people on social media but ends up adopting the bad side of the conversation as they interact [16]. This serves as an example of the need for guidance on the purpose of the chatbot we are creating. There have been many positive stories about and use of chatbots, but the question remains will they create a society of its own in the already disconnected world of social media or will they help to shape up the behaviour of the humans which to some extent has already degraded.

## VI. ARTIFICIAL STUPIDITY

With the unpredictable behaviour of humans, learning from them can sometimes be difficult. The data that AI technology is using to train and learn mostly does not cover all the aspect of human behaviour. Intelligence comes from learning, whether you're human or machine. Systems usually have a training phase in which they "learn" to detect the right patterns and act according to their input. Once a system is fully trained, it can then go into the test phase, where it is hit with more examples and we see how it performs. The training phase cannot cover all possible examples that a system may deal with in the real world. These systems can be fooled in ways that humans wouldn't be. For example, random dot patterns can lead a machine to "see" things that aren't there. If we rely on AI to bring us into a new world of labour, security, and efficiency, we need to ensure that the machine performs as planned and that people can't overpower it to use it for their ends.

---
[1] https://visabot.co/

## VII. AI BIAS

For a human to be intelligent they need to learn to do the AI systems. But the imperfect nature and of humans creates a biased environment when it comes to the dataset the models are learning from. One good example could be a criminology prediction system which was observed to be biased against black people [17]. Although the model proved to be accurate still the assessment of the prediction together with the errors produced was not assessed enough. Here comes the question of the purpose and the impact of technology on people's lives. The analysis and handling of the technology need in-depth exploration of the results it produced and not only rely on the accuracy. Handling these systems to a non-expert to rely on can be more detrimental than it looks. Although the systems are building to serve our social progress the bias nature of humanity can still be propagated if not handled carefully. In another research conducted by the Massachusetts Institute of Technology together with Stanford University were the team examined the three facial recognition systems and broke down the accuracy of the result based on gender and race [18]. The researchers noted biases in the results produced and noted that that classification bias came from the dataset used to train the model. So, the researchers in AI argue that the biases in the kind of system we produced can be reduced by diversifying the workforce that is involved in building the systems or even making the researchers aware of the biases that exist in their work.

## VIII. SECURITY

Autonomous weapons are now being developed at a rapid pace. Powerful nations are battling each other to become the leading power in autonomous weapons. From automated warfare jets, to robotic soldiers all are already in the ground field. The increased power of destruction that the technology is adding is far beyond measures. Talking about nuclear, aerospace, cybersecurity, and biotechnology these are priority areas for the country's national security. For example, Cybersecurity is becoming even more important now as the fight won't only be on the battleground but the systems we are building ourselves. Big countries like the US, China, and Russia are racing each other towards the global dominance of AI technology. A quote from the Russian president Vladimir Putin said, "AI is the future, not only for Russia but for all humankind," Whoever becomes the leader in this sphere will become the ruler of the world." [19]. Countries like China have even announced its ambition of becoming the leader in AI technology by the year 2030 [20]. This shows the race is on and only time can tell the outcome of the system that is faster and more capable than humans by order of magnitude.

## IX. EVIL GENES

With many unanswered questions within the AI field today, new questions and unknowns continue to erupt as the technology progresses. The evil within the AI is not what we see in Hollywood movies of machines turning against humans, but rather is the mechanisms that the AI uses to solve the problems which we did not intentionally do that way. An interesting example is provided by Julia Bossmann in her report where an AI system could provide an answer to eliminate all humans as a solution to combat cancer [21]. So as useful as it is the technology still needs to understand the full context of the environment in which we want to find a solution for that.

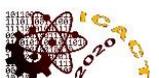

## X. THE ECONOMIC IMPACT OF AI

Like in the early days when electricity was discovered and brings about the industrial revolution, the same is predicted for AI technology to bring another industrial revolution. In an article published by Parkes and Wellman, the four sectors that will see the tremendous effect of AI technology are the Manufacturing industries, Professional services, financial services, and wholesale and retail [22]. Some countries are already starting to take advantage of the opportunities presented by technology today. One of the countries is China which has even set its goal to become the global leader in AI research by 2030 [20]. With a massive population of around 1.4 billion in which 730 active internet users, the country sees its potential in collecting data, and the technology becoming more hungry for data. It seems that AI is going to become the new powerhouse for the countries and investment in research will be crucial for the country to break through and beneficial utilization of technology.

According to research conducted by Accenture, one of the leading global technology companies on top twelve global economic leaders revealed that AI technology could double annual economic growth rates in 2035 by changing the nature of the work and creating new relationships between man and machine [23]. This indicates an increase in the absorption of AI in our economies. The technology will continue to offer amplification and transcend of the current capital and labour capacity to propel our economic growth. Although the initial phase will be more on the industrial and manufacturing side, consumers will be the next frontier as a variety of AI products will finally be on the market.

## XI. CONCLUSION

With this exponential increase in AI technology, it is time for the people and their governments to embrace technology instead of fearing change. It is also clear that the complication of the challenges that technology presents is also a concern. Governments and other policymakers should collaborate with the researchers to try and steer the technology towards serving humanity as the impact of the technology spans globally. As the computation power advances and more data being generated, researchers together with the community should focus on making both technology and people working more together rather than increasing the gap. Different organizations, associations, and individuals are becoming global examples in collaborating to take action and measure against the challenges that the technology presents. Organizations such as Algorithmic Justice League which highlights the algorithmic bias and let the people raise their concerns and experience to develop best practice and accountability. The petition to the UN urging rapid action on weaponized AI will help global engagement. The AI4All which aims to train a new and more diverse generation of the future AI technologist, thinkers, and leaders [3]. The is a need for global engagement for AI to stimulate and enforce policymakers to take appropriate actions.